\documentclass[aps,prl,twocolumn,showpacs,preprintnumbers,amsmath,amssymb,superscriptaddress]{revtex4-2}
\usepackage{graphicx}% Include figure files
\usepackage{dcolumn}% Align table columns on decimal point
\usepackage{bm}% bold math
\usepackage{xcolor}
\usepackage{txfonts}
\usepackage{hyperref}
\hypersetup{colorlinks,citecolor=red,filecolor=black,linkcolor=blue,urlcolor=blue}
\usepackage{natbib}
\usepackage[utf8]{inputenc}
\usepackage{calc}
\usepackage{accents}

\newcommand{\nc}{\newcommand}
\nc{\be}{\begin{equation}}
\nc{\ee}{\end{equation}}
\nc{\bea}{\begin{eqnarray}}
\nc{\eea}{\end{eqnarray}}
\nc{\bean}{\begin{eqnarray*}}
\nc{\eean}{\end{eqnarray*}}
\nc{\mb}{\mbox}
\nc{\rnc}{\renewcommand}
\nc{\vk}{\mb{\bf k}}
\nc{\vp}{\mb{\bf p}}
\nc{\vn}{\mb{\bf n}}
\nc{\vq}{\mb{\bf q}}
\nc{\rr}{\mb{\bf r}}
\nc{\vz}{\hat {\mb{\bf z}}}
\nc{\vj}{\mb{\boldmath$j$}}
\nc{\vg}{\mb{\boldmath$g$}}
\nc{\x}{\mb{\boldmath$x$}}
\nc{\A}{\mb{\boldmath$A$}}
\nc{\va}{\mb{\boldmath$a$}}
\nc{\vs}{\mb{\boldmath$\sigma$}}
\nc{\vpi}{\mb{\boldmath$\pi$}}
\nc{\nab}{\nabla}
\nc{\X}{\sf x}
\nc{\kk}{{\bf k}}
\nc{\pp}{{\bf p}}
\nc{\qq}{{\bf q}}
\nc{\Qq}{{\bf Q}}
\nc{\Rr}{{\bf R}}
\nc{\vl}{{\bf l}}
\nc{\Kk}{{\bf K}}
\nc{\upspin}{{\uparrow}}
\nc{\dspin}{{\downarrow}}
\nc{\vecq}{{\bf q}}
\nc{\veck}{{\bf k}}
\nc{\vecp}{{\bf p}}
\nc{\vecl}{{\bf l}}
\nc{\vecr}{{\bf r}}
\nc{\vecx}{{\bf x}}
\nc{\vecG}{{\bf G}}
\nc{\vecA}{{\bf A}}
\nc{\vecpi}{{\bf \pi}}
\nc{\vecL}{{\bf L}}
\nc{\vecK}{{\bf K}}

\nc{\im}{\imath}

\nc{\argg}{\text{Arg}}
\nc{\bd}{\textbf}
\nc{\bds}{\boldsymbol}
\nc{\ham}{\hat{\mathcal{H}}}
\nc{\la}{\langle}
\nc{\ra}{\rangle}
\nc{\re}{\text{Re}}
\nc{\rn}[1]{%
	\textup{\uppercase\expandafter{\romannumeral#1}}%
}
\nc{\sgn}{\text{Sgn}}
\nc{\tit}{\textit}
\nc{\tr}{\text{Tr}}
\nc{\les}{\leqslant}
\nc{\ges}{\geqslant}

\usepackage{stackengine}

\begin{document}

\title{Kekul\'{e} Superconductivity Induced by Quantum Geometry in Haldane Phases}

\author{Yafis Barlas}
\email{ybarlas@unr.edu}
\affiliation{Department of Physics, University of Nevada, Reno, Nevada 89557, USA}
\author{Fan Zhang}
\affiliation{Department of Physics, University of Texas at Dallas, Richardson, Texas 75080, USA}
\author{Enrico Rossi}
\affiliation{Department of Physics, William and Mary, Williamsburg, Virginia 21387, USA}

\begin{abstract}
Chiral two-dimensional electron gases, which capture the electronic properties of graphene and rhombohedral graphene systems, exhibit singular momentum-space vortices and are susceptible to interaction-induced topological Haldane phases. Here, we investigate pairing interactions in the inversion-symmetric Haldane phases
of chiral two-dimensional electron gases. We demonstrate that the nontrivial band topology of the Haldane phases enhances intra-valley ($\Qq = \pm 2 \Kk_D$) pair susceptibility relative to inter-valley ($\Qq = 0$) pair susceptibility, favoring the emergence of a lattice-scale pair-density wave order.  When longitudinal acoustic phonons mediate the pairing interaction, the system supports a chiral Kekul\`{e} superconducting order at low densities. At higher densities, the phase diagram depends on the parity of the Chern number $J$. For even parity, an s-wave Kekul\`{e} order appears, while for odd parity we find a $\Qq =0$ valley triplet chiral-$J$ and a valley singlet s-wave order at different densities. Our findings are relevant to superconductivity in rhombohedral graphene and Kagome metals.
\end{abstract}

\maketitle
 
While significant progress has been made in understanding unconventional 
superconductivity~\cite{Cao2018,Yankowitz1059,Cao2018second,SCMAtTLGPablo,TripletSCtTLGPablo,Lin2022,Scammell2022,Xia2025,Guo2025,Burg2022,BLGsuperconductivity} in spin Chern bands of twisted two-dimensional (2D) crystals, the presence of a moir\'e
superlattice often complicates the identification of the underlying mechanisms~\cite{PhysRevB.98.220504,PhysRevX.8.031089,PhysRevLett.121.087001,PhysRevB.98.075154,PhysRevB.98.121406,PhysRevLett.121.217001,PhysRevX.8.041041,PhysRevLett.122.026801,PhysRevB.100.205113,PhysRevLett.127.217001,PhysRevB.104.L121116,PhysRevLett.121.257001,PhysRevB.99.165112,PhysRevLett.122.257002,PhysRevB.98.241412,PhysRevB.98.054515,GuineaPNAS,SkyrmionSC}. For instance, it remains unresolved whether the observed superconductivity originates predominantly from band-projected interactions or from novel pairing instabilities intrinsic to topological bands~\cite{Han2024,Lu2024,ABCrGSOC,ABCAstackQAH,Lu2025}. 
 Alternatively, multi-layer rhombohedral graphene (RG) has emerged as a promising platform, displaying superconducting phenomenology~\cite{Jahin2024a,Kim2025a,Chou2024a,Geier2024a,Yang2024b,Wang2024a,Qin2024a,Yoon2025a,parramartinez2025} 
that appears to be as rich as that of twisted 2D crystals~\cite{Zhou2021a,Han2025,kumar2025superconductivitydualsurfacecarriersrhombohedral,morissette2025stripedsuperconductorrhombohedralhexalayer}, but without the complexity of a moir\`{e} potential. Notably, in RG, superconductivity can even be hosted by a spin-polarized valley
Chern band~\cite{Han2025,Jahin2024a,Kim2025a,Chou2024a,Geier2024a,Yang2024b,Wang2024a,Qin2024a,Yoon2025a,parramartinez2025,wy3f-hgr9},
which underscores the necessity to understand the interplay of band geometry and pairing interactions~\cite{JiangBarlasPDW}.
 
In ordinary metals, the pair susceptibility typically diverges at zero center-of-mass momentum ($\Qq = 0$), favoring uniform superconductivity over finite-momentum pairing at weak coupling~\cite{Mahan_book,ketterson_song_1999,coleman_2015}. In contrast, this Letter demonstrates that the nontrivial topology of a Chern band qualitatively modifies this picture. Previous works have shown that the quantum geometry can strongly affect the response of superconductors to electromagnetic fields, especially in flat-band systems~\cite{ROSSI2021100952,Torma2022,Yu2025}. 
Our results demonstrate a very different key role of quantum geometry. We show that even for
standard effective attractive interactions, such as those mediated by acoustic phonons, the 
quantum geometry of the Bloch states at the Fermi energy can lead to unconventional
superconducting states characterized by a finite center-of mass momentum and a chiral order parameter.

 Our conclusions are based on the analysis of superconducting pairing within a Chern band realized in chiral two-dimensional electron gases (C2DEGs). Specifically, we show that in a spin-polarized half-metal phase of RG, the band-projected intra-valley pair susceptibility with pair momentum $\Qq = \pm 2 \Kk_D$ is enhanced relative to the inter-valley pair susceptibility with $\Qq=0$, and depends on the ratio $m/\mu$, where $\mu$ is the Fermi energy 
and $m$ is an interaction-induced Haldane mass~\cite{PhysRevLett.106.156801}. Since this enhancement occurs at high-symmetry points in the Brillouin zone, it naturally favors a lattice-scale pair density wave (PDW) order, previously identified as a Kekul\'{e} superconductor in a different context~\cite{PhysRevB.82.035429}.

\begin{figure}
\begin{center}
\includegraphics[width=0.48\textwidth,clip]{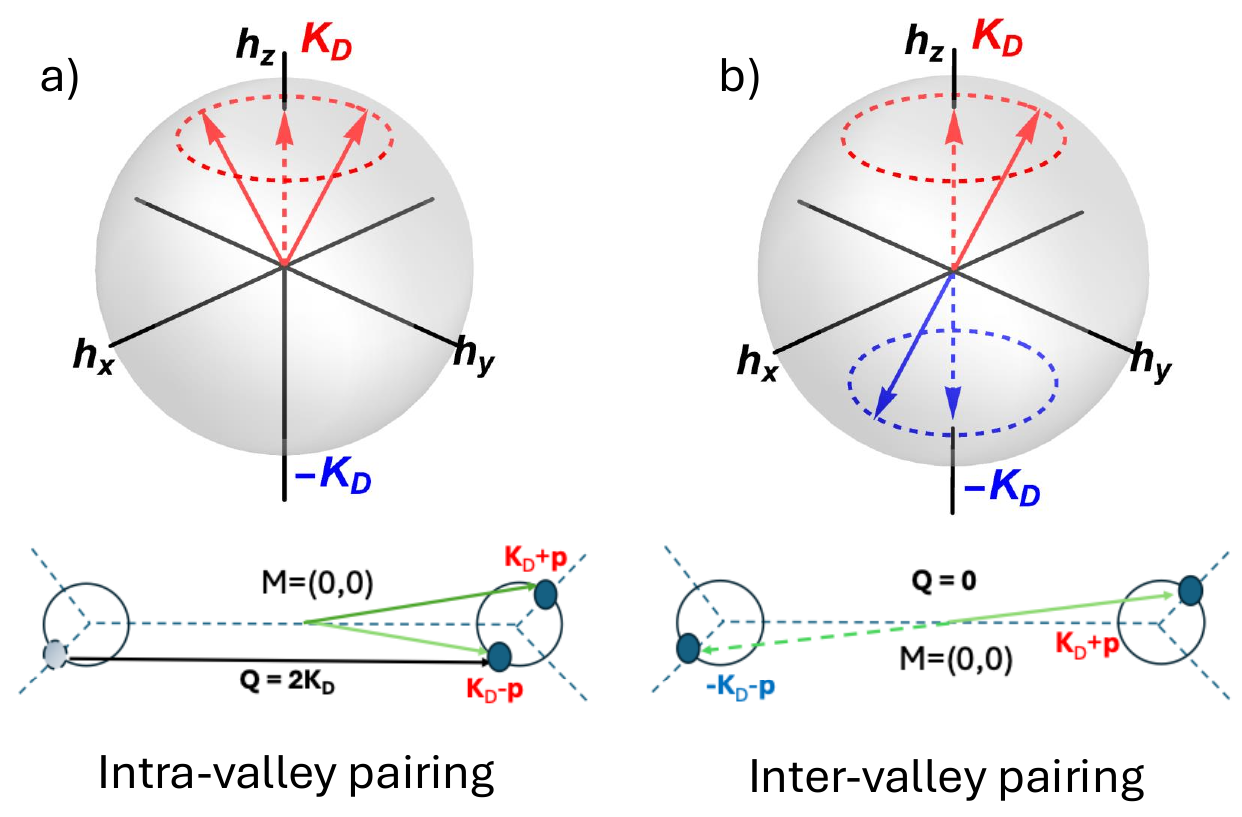}	
\caption{Hamiltonian pseudo-spinor field $\hat{{\bf h}}$ on the Bloch sphere for a) intra-valley pairing and b) inter-valley pairing in the 
Haldane phase, with the Fermi surface projection indicated by the dotted red (blue) circles for $\Kk_D\,(-\Kk_D)$ valleys. 
The projections of $\hat{{\bf h}}$ onto the $\hat{z}$-direction for intra-valley pairs $\pp$ and $-\pp$ are parallel, whereas they are 
anti-parallel for inter-valley pairs $\pp$ and $-\pp$. This spinor structure of the Haldane phase results in the suppression of the coherence factors of the inter-valley pair susceptibility (see text for details).}
	\label{fig:Pairsusc}
	\end{center}
\end{figure}

This enhancement arises from the coherence factors of the Bloch wavefunctions in the Haldane phase~\cite{footnote1}, which can be visualized using the $\hat{z}$-component of the pseudo-spinor field defined by the Hamiltonian on the Bloch sphere, as illustrated in Fig.~\ref{fig:Pairsusc}. In the case of intra-valley pairing, as depicted in Fig.~\ref{fig:Pairsusc} a), the $\hat{z}$-component pseudo-spinors point in the same direction, resulting in constructive interference. In contrast, for inter-valley pairing shown in Fig.~\ref{fig:Pairsusc} b), the pseudo-spinors point in opposite directions, leading to destructive interference and a corresponding suppression of the inter-valley pair susceptibility. Overall, the pair susceptibility and superconducting phase diagram depend on the parity of the Chern number $J$. When longitudinal acoustic phonons mediate the pairing interaction, the system supports a chiral Kekul\`{e} superconducting order for low densities $n_e$. For even values of $J$ the system transitions to an s-wave Kekul\`{e} superconducting order when $n_e > n_{e,2}$, with a chiral Kekul\`{e} superconducting order appears for $n_e < n_{e,2}$. For odd values of $J$, a $\pi$-phase chiral Kekul\`{e} superconducting order appears when $n_e < n_{e,1}$, while in the density regime $n_{e,1}<n_e<n_{e,2}$, we find a $\Qq=0$ valley triplet chiral superconducting state and a valley-singlet s-wave superconductor for $n_e > n_{e,2}$. The chiral superconducting order arises from projected interactions within the topological bands, thereby further reflecting the underlying geometric features of the Haldane phase of C2DEGs. 

% We discuss the properties of the chiral Kekul\'{e} superconductors and the relevance of our findings to multi-layer RG.

{\em Model}.---The continuum model Hamiltonian of a C2DEG with a Chern number $J$~\cite{footnotedef} on a projected bipartite lattice is 
\be
\label{eq:Hamchiral} 
H_0 = \zeta_J p^J \bigg( \hat{\sigma}_x \cos(J \varphi_{\pp}) + \tau_z \hat{\sigma}_y \sin(J \varphi_{\pp}) \bigg) + m \hat{\sigma}_z \tau_z ,
\ee
where the Pauli matrices $\hat{\sigma}_{i}$ act on the pseudospinor $\psi^{\dagger} =(\psi^{\dagger}_{A},\psi^{\dagger}_{B})$, $(A, B)$ are sublattice degrees of freedom to be specified below, $\tau_z=\pm 1$ denote the valleys associated with the Dirac points located at the corners of the hexagonal Brillouin zone (BZ) $\pm\Kk_D$, $\pp =(p_x,p_y)$ is the momentum measured from the Dirac points with $\varphi_{\pp} =\tan^{-1}( p_y/p_x)$, and $p  = \sqrt{p_x^2 +p_y^2}$. We choose the $M$-point of the BZ as the origin, giving $\pm\Kk_D=(\pm2 \pi/3,0)$ and the valley separation $\Qq=2\Kk_D = (4 \pi/3,0)$. Eq.~\ref{eq:Hamchiral} with $m=0$ describes the minimal model of $J$-layer RG~\cite{PhysRevB.77.155416}, with the pseudospinor defined on the top $(1,\,A)$ and bottom $(J,\,B)$ layers, $\zeta_{J} = v^J/\gamma_1^{J-1}$, and $v= \sqrt{3}/2 \gamma_0 a_0 \sim 673$~meV nm$^{-1}$, where $\gamma_0 \sim 3.16$~eV and $\gamma_1\sim 0.445$~eV are the nearest-neighbor intra- and inter-layer hopping parameters~\cite{Yoon2025a}, respectively, and $ a = 0.246 $~nm is the graphene lattice constant. To highlight the interplay between band geometry and pairing interactions, we neglect trigonal warping for now and address this aspect later. 

The Haldane mass term $ m\hat {\sigma}_ z\tau_z$ results in a quantum anomalous Hall 
(QAH) state with a Chern number $J$~\cite{PhysRevLett.106.156801}. 
This mass term breaks time-reversal symmetry while preserving inversion symmetry.
In contrast, a displacement field produces a sublattice staggered potential $\Delta_V\hat\sigma_z$ breaking the inversion symmetry. The band dispersion $\epsilon_{\pp,J,\pm} = \pm  (\zeta_J^2 p^{2J} + m^2)^{1/2}$ near the $\pm\Kk_D $
points is particle-hole and inversion symmetric. The Bloch wave function for the conduction band near $\Kk_D$ is $| u_{\pp}( \Kk_D) \rangle = (\cos (\theta_{\pp}/2)e^{-\im \tau_z J \varphi_{\pp}/2},\sin (\theta_{\pp}/2)e^{\im  \tau_z J \varphi_{\pp}/2} )$ with $\cos(\theta_{\pp}) = m/\epsilon_{\pp,J,+}$, with inversion symmetry giving $| u_{-\pp}(- \Kk_D) \rangle = \sigma_{x} | u_{\pp}( \Kk_D) \rangle $. The singularities of the Bloch states near the Dirac points exhibit frustrated momentum-space pseudospin structure, creating a ``fertile ground" for spontaneous symmetry breaking~\cite{PhysRevB.77.041407,PhysRevB.81.041402,PhysRevLett.106.156801}. 

{\em Interaction-induced Haldane phase in RG}.---Since our focus is on the superconducting instabilities of the Haldane phase,  we consider the simplest case of spinless C2DEGs. In spinless C2DEGs, the corresponding valley-diagonal mass terms are i) $m \hat{\sigma}_z \tau_z$ and ii) $\Delta_{V} \hat{\sigma}_z$. The latter leads to a valley Hall effect, 
because time-reversal symmetry ensures that the Berry curvatures at the two valleys are equal in magnitude but opposite in sign. In contrast, the former induces a QAH effect with $\sigma_{xy} =J e^2/h$, since inversion symmetry dictates that the Berry curvatures at the two valleys are identical, resulting in a nontrivial Chern number.

A series of experiments~\cite{Geisenhof2021,ABCrGSOC,ABCAstackQAH,Winterer2024} have provided compelling evidence for interaction-driven Haldane phase in RG systems ranging from bilayer to pentalayer. 
In particular, the QAH conductance is quantized at $5e^2/h$ in pentalayer RG at zero magnetic field~\cite{ABCrGSOC}. 
Effective in these states, only one spin contributes to the QAH effect~\cite{PhysRevLett.106.156801}.
Our mean-field analysis for spinless C2DEGs (see End Matter) reveals that a short-ranged valley-dependent negative interaction $u_{\perp} < 0$ in the ${\bm \tau}_{\perp} \equiv {\bm \tau} \times \tau_z $ channel leads to a spontaneous QAH state, whereas $u_{\perp} > 0$ yields a fully-layer-polarized state. The interaction-induced Haldane mass term $m$ depends on $J$, which can be significant: $m \sim 50$  meV for $J=4,5$ for a hBN-encapsulated device with a dielectric constant $\epsilon \sim 5$, for gate-screened interactions with gate distance $d= 10$ nm. Since mean-field theory tends to overestimate interaction-induced gaps, we take $m \neq 0$ as a tunable parameter.

\begin{figure}
\begin{center}
    \includegraphics[width=0.45\textwidth,clip]{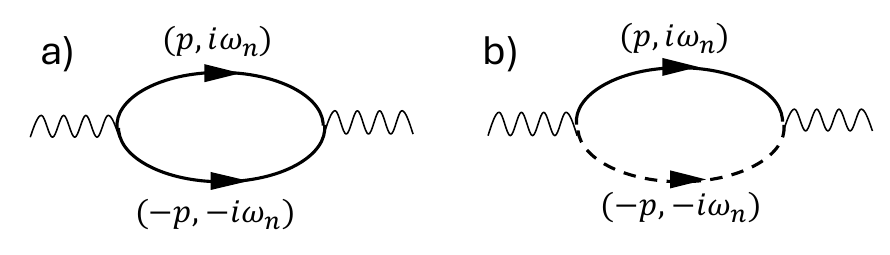}	
	\caption{Feynman diagrams for the a) intra- and b) inter-valley pair susceptibility of chiral 2DEGs. The solid (dotted) lines denote the band-projected electron propagator of the $\Kk_D(-\Kk_D)$ valleys. }
	\label{fig:FeyndiagPP}
	\end{center}
\end{figure}

{\em Pair susceptibility in the Haldane phase}.---One key finding of this Letter is that the pair susceptibility of the Haldane Chern bands is strongly governed by the topologically enforced geometrical properties of the Bloch states, leading to qualitatively distinct superconductivity compared to that of trivial bands. For a generic two-orbital Hamiltonian of the form $H_0 = {\hat{\bm\sigma}} \cdot {\bf h}(\kk)$, where ${\bf h}(\kk)$ defines the momentum-dependent Hamiltonian field, the projected pair susceptibility $\Pi (\qq)$ in the $\sigma_0$-channel (see Fig.~\ref{fig:FeyndiagPP}) can be expressed as (see Supplemental Materials~\cite{supplement1})
\be
\label{eq:Pairsusc}
\Pi(\qq) = \int \frac{d^2 \kk}{(2 \pi)^2} \frac{g(\kk,\kk+\qq)}{\xi_{\kk+\qq}+ \xi_{-\kk}} \bigg( \frac{1+ \hat{\bf h}_{\kk+\qq} \cdot \hat{\bf h}_{-\kk}}{2} \bigg) , 
\ee
where $\hat{\bf h}_{\kk} = ( \hat{h}_x(\kk), \hat{h}_y(\kk), \hat{h}_z(\kk))$ is the normalized Hamiltonian vector field $\hat{\bf h}_{\kk} = {\bf h}_{\kk}/|{\bf h}_{\kk}|$, $\xi_{\kk} = \epsilon_{\kk} - \mu$ denotes the energy measured from the Fermi surface, and $ g(\kk,\kk+\qq)= 1 - n_{F} (\xi_{\kk+\qq}) - n_{F} (\xi_{-\kk})$, $n_F(\epsilon) = (e^{\beta \epsilon}+1)^{-1}$ is the Fermi-Dirac distribution with $\beta =1/(k_{B}T)$, and $\kk \in$ BZ.

As illustrated in Fig.~\ref{fig:Pairsusc}, $\qq =0+\pp$ corresponds to the inter-valley pair susceptibility, whereas $\qq = 2 \Kk_D + \pp$ denotes the intra-valley pair susceptibility. Due to the topological nature of the band $\hat{\bf h}_{\Kk_D} \sim (0,0,1)$ and $\hat{\bf h}_{-\Kk_D} \sim (0,0,-1)$. Therefore, the coherence factors of inter-valley pair susceptibility vanish as $(1+ \hat{\bf h}_{\pp} \cdot \hat{\bf h}_{-\pp})/2 \to 0$, while the intra-valley pair susceptibility is unaffected, $(1+ \hat{\bf h}_{\pp} \cdot \hat{\bf h}_{-\pp+2 \Kk_D})/2 \to 1$ (see Fig.~\ref{fig:Pairsusc}). Thus, intra-valley pairing remains robust near the $\pm\Kk_D $, whereas inter-valley pairing is suppressed due to topologically enforced destructive interference of band geometry. 

Away from the Dirac points, the intra-valley (S) and inter-valley (D) pair susceptibilities, denoted by $\Pi_{S}= \Pi(\Qq= 2\Kk_D)$ and $\Pi_{D} = \Pi(\Qq=0)$, depend on the ratio of the Haldane gap $m$ to the Fermi energy $\mu$, Chern number parity, and can be expressed as 
\be 
\label{eq:pairsusceptibility}
%\Pi_{S} =-\mathcal{D}(\mu) \log\bigg( \frac{E_c}{T} \bigg)  f_{D}(x)+ \cdots, \\ 
\Pi_{a} = \frac{N_0}{(1-x^2)^{\alpha}} \log\bigg( \frac{E_c}{T} \bigg)  f_{a}(x)+ \cdots,
%\Pi_{D} = \frac{N_0}{(1-x^2)^{\alpha}} \log\bigg( \frac{E_c}{T} \bigg)  f_{S}(x) + \cdots, 
\ee    
where  $\cdots$ denotes non-singular terms, $a=S(D)$ denotes the intra- and inter-valley pairing, $N_0 = x^{1-\beta}/(2 \pi J \zeta_J^\beta m^{1-\beta}) $, denotes the characteristic density of states, with $\alpha = 1-1/J $, $\beta = 2/J$, $E_c \sim \gamma_1$ is the high-energy cutoff, and we have set the Boltzmann constant $k_{B} =1$. The intra-valley (S) and inter-valley (D) pair susceptibilities depend on $x= m/\mu$ through the functions $f_{S}(x)$ and $f_{D}(x)$.  For even Chern numbers $f_{S}(x)=1$ and $f_{D}(x) =(1-x^2)/2$, while for odd Chern numbers $f_{S}(x) = x^2$ and $f_{D}(x) =(1-x^2)/2$. Therefore, for even values of $J$, $\Pi_{S} > \Pi_{D}$ throughout the physically relevant range $x \in (0,1]$. Alternatively, for odd values of $J$, $\Pi_{S} > \Pi_{D}$ requires $x > 1/\sqrt{3}$. In both cases, the geometric suppression of inter-valley coherence represents a hallmark of pairing in the Haldane Chern bands. 

\begin{figure}
\centering
 \includegraphics[width=0.45\textwidth,clip]{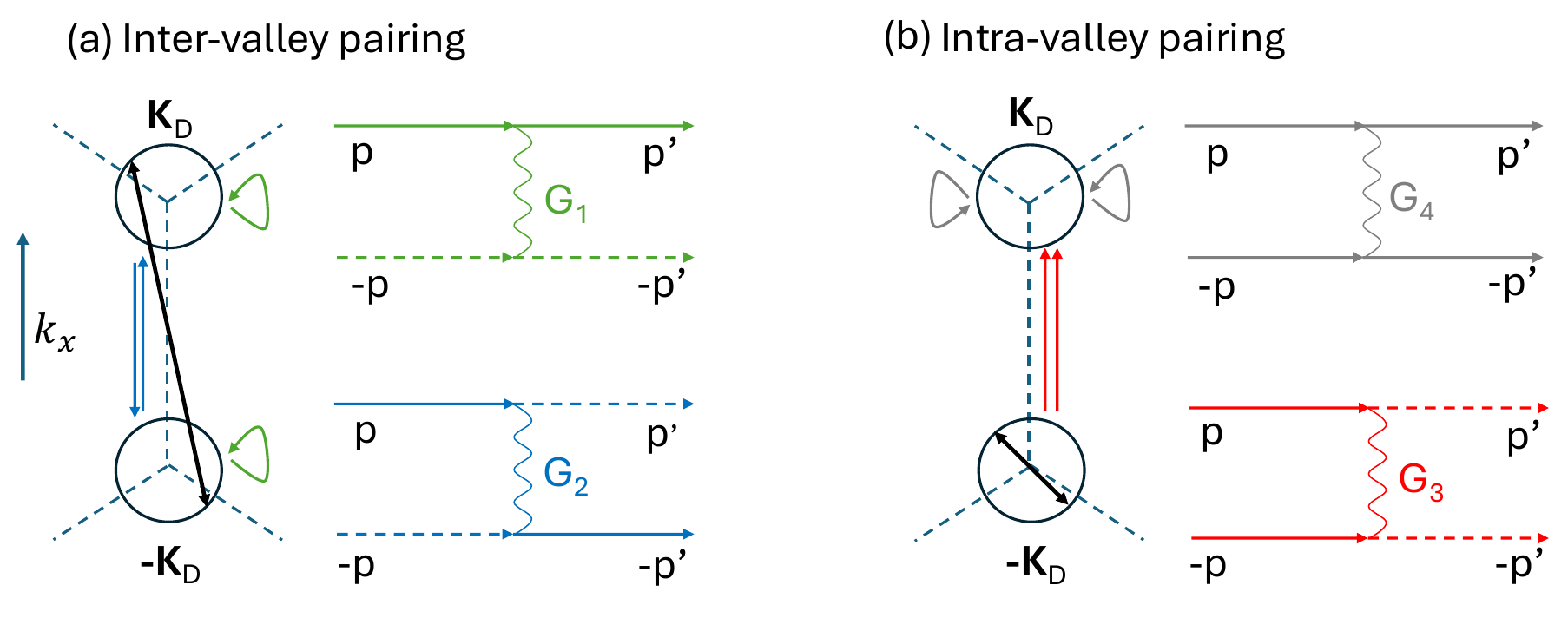}	
	\caption{a) Feynman diagrams for inter-valley scattering interaction $G_1$ and inter-valley exchange interaction $G_2$, and b) Feynman diagrams for intra-valley scattering interaction $G_4$ and pair tunneling $G_3$. The solid (dotted) lines correspond to the electron propagator in the valley $\Kk_D$ $(-\Kk_D)$. The colors correspond to the microscopic interactions on the circular Fermi surfaces for both inter-valley and intra-valley pairing.}
	\label{fig:LAinteractionChern}
\end{figure}

{\em Longitudinal acoustic phonon-mediated pairing}.---To determine whether the enhanced intra-valley pair susceptibility leads to an intra-valley superconducting state, we must specify a model for the pairing interaction. Our estimates indicate that longitudinal acoustic (LA) phonons provide the leading contribution to the effective pairing interaction for a circular Fermi surface (see Supplemental Materials~\cite{supplement2}). Consequently, we neglect the contribution of optical phonons in our analysis. We note, however, that optical phonons correspond to an off-diagonal electron-phonon coupling in orbital space~\cite{PhysRevLett.121.257001,PhysRevB.110.045133} and therefore favor inter-valley pairing (see End Matter).

Since translational symmetry requires that the interaction is explicitly independent of $\Qq$, any $\Qq$ dependence of the pairing interaction must arise from the projection onto the Haldane Chern bands. By employing a continuum model for the electron-phonon coupling and performing a Schrieffer–Wolff transformation on the band-projected interaction, we arrive at the generic form of the effective pairing interaction (see End Matter) with two Fermi surfaces at different valleys,
\bea
\label{eq:HamSSHpair}
\nonumber
H_{pair} &=& \sum_{\pp,\pp'} \bigg[ G_1(\pp,\pp') a^{\dagger}_{\pp} b^{\dagger}_{ -\pp} b_{-\pp'} a_{\pp'} + 
G_2(\pp,\pp') a^{\dagger}_{\pp} b^{\dagger}_{ -\pp} a_{-\pp'}   b_{\pp'} \\ 
\nonumber
&+& G_3(\pp,\pp') a^{\dagger}_{\pp} a^{\dagger}_{ -\pp} b_{-\pp'}  b_{\pp'} + G_4(\pp,\pp') a^{\dagger}_{\pp} a^{\dagger}_{ -\pp} a_{-\pp'} a_{\pp'} \\ 
&+& (a \leftrightarrow b) \bigg],
\eea
where the electron creation operators at $\Kk_D(-\Kk_D) $ are defined as $c^{\dagger}_{\Kk_D+\pp} = a^{\dagger}_{\pp} $ and $c^{\dagger}_{-\Kk_D+\pp,} = b^{\dagger}_{\pp} $. Due to inversion symmetry, the interaction matrix elements $G_i(\pp,\pp')$, satisfy $G_{1}(\pp,\pp') = 2 G_{4}(\pp,\pp')$ and $G_{2}(\pp,\pp') = 2G_{3}(\pp,\pp')$. Near the Fermi surface $p_x = p_F \cos(\varphi_{\pp}), p_y = p_F \sin(\varphi_{\pp}) $ and $|\pp| \sim |\pp'| \sim p_F$, the interaction matrix elements can be expressed as, 
\bea
\nonumber
G_{4} &=& -G_0 e^{- \im J\Phi} \bigg( 
\frac{e^{\im J\varphi_{\pp}}(\xi_p + \mu - m )}{2(\xi_p + \mu)} + \frac{e^{\im J\varphi_{\pp'}}(\xi_p + \mu + m )}{2(\xi_p + \mu)} \bigg)^2, \\
G_{3} &=&  -\frac{G_0}{2} \bigg[ 1+ (-1)^J \cos ( J \Phi) \bigg]  \bigg( 1 - \frac{m^2}{(\xi_p + \mu)^2} \bigg),
\eea
where $\Phi = \varphi_{\pp} + \varphi_{\pp'}$ and the pairing interaction $G_0 = 9.24 \times 10^{-8} (D/(\hbar v_s))^2 \sim 123 $~meV$\cdot$nm$^{2}$, with $D \sim 15$~eV denoting the deformation field~\cite{footnote2}, and $v_s \sim 2 \times 10^4 $~m/s is the phonon sound velocity~\cite{PhysRevLett.127.187001,PhysRevB.99.165112}. In Eq.~\ref{eq:HamSSHpair}, the interactions $G_1$ and $G_2$, which denote inter-valley scattering and inter-valley exchange, as indicated in Fig.~\ref{fig:LAinteractionChern}~a), naturally lead to a conventional superconducting state. Intra-valley pairing results from $G_4$ and $G_3$, which label intra-valley scattering and pair tunneling between the two valleys, as shown in Fig.~\ref{fig:LAinteractionChern}~b). 

To determine the leading pairing instability, we decompose each interaction $G_{i,l}$ into the angular momentum channels $l=0,J,-J$. The renormalization group flow equations (see Supplemental Materials~\cite{supplement3}) for the pairing interactions $G_{i,l}$ can be expressed as
\bea
\label{eq:RGinter}
\dot{g}_{1,l} &=& -[g_{1,l}^2 + g_{2,l}^2]f_{D}(x), \quad \dot{g}_{2,l} = -2 f_D(x) g_{1,l} g_{2,l}, \\
\label{eq:RGintra}
\dot{g}_{4,l} &=& -[g_{4,l}^2 + g_{3,l}^2] f_{S}(x), \quad \dot{g}_{3,l} = -2 f_S(x) g_{3,l} g_{4,l},
\eea
where $g_{i,l}(x) = N_0 G_{i,l}/(1-x^2)^{\alpha}$ and $\dot{g_{i,l}} = dg_{i,l}/dt$, where $ t = \log(\omega_D/T)$ and $\omega_D \sim 0.2$ eV denotes the Debye energy. Eq.~\ref{eq:RGinter} and Eq.~\ref{eq:RGintra} can be solved by making the substitution $g^{intra}_{\pm,l} = g_{4,l} \pm g_{3,l}$, $g^{inter}_{\pm,l} = g_{1,l} \pm g_{2,l}$ which gives 
\be 
\label{eq:RGsolution}
g^{intra}_{\pm,l} = \frac{g^{intra}_{\pm,l}}{1+f_{S}(x) g^{intra}_{\pm,l}  t}, \quad g^{inter}_{\pm,l} = \frac{g^{inter}_{\pm,l}}{1+f_{D}(x)g^{inter}_{\pm,l} t}.
\ee
The superconducting critical temperature for each pairing channel is determined from the poles of Eq.~\ref{eq:RGsolution}, and depends on the parity of the C2DEG, as discussed below.

\begin{figure}
\begin{center}
    \includegraphics[width=0.48\textwidth,clip]{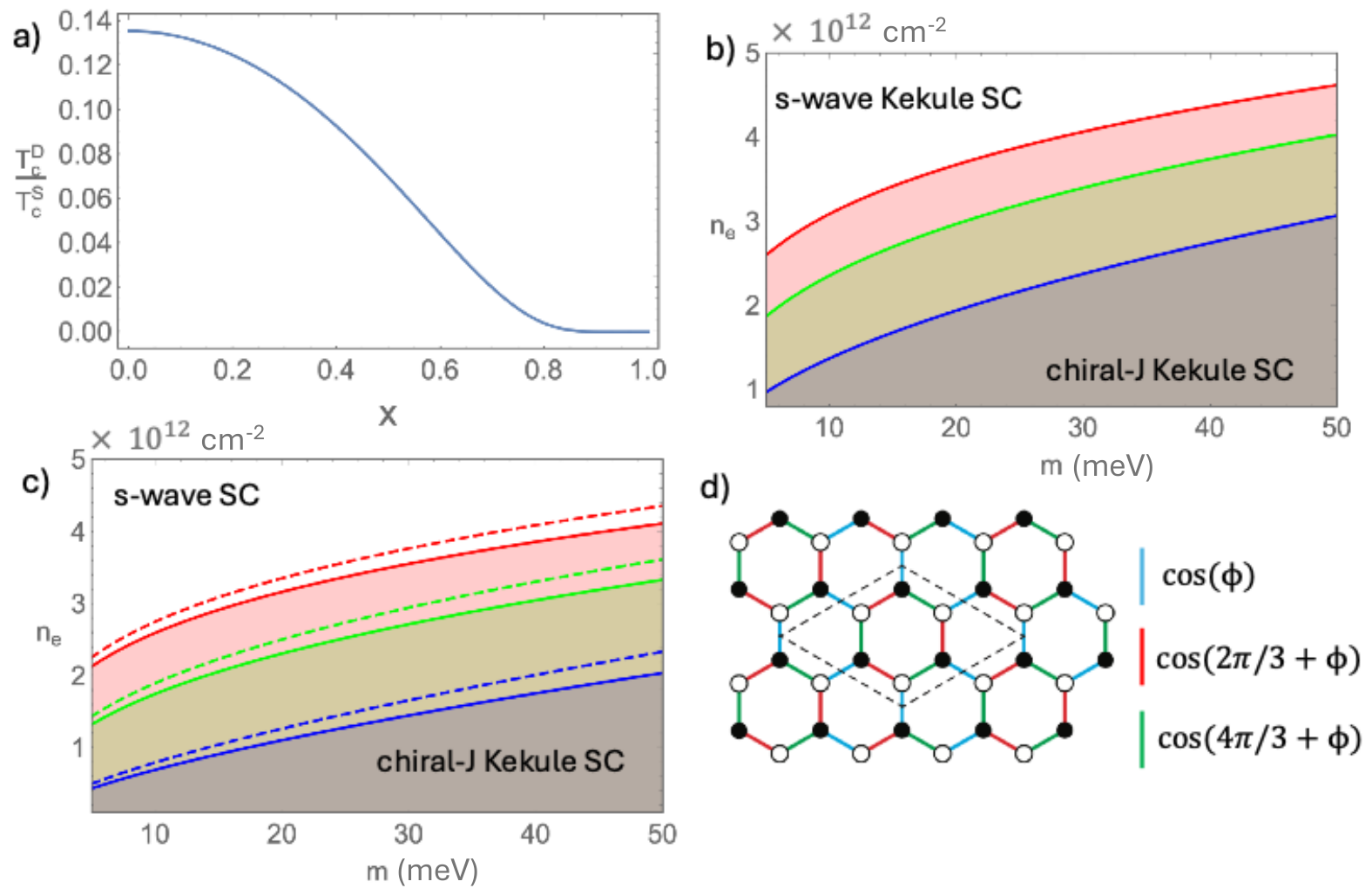}	
	\caption{a) The ratio of the inter-valley critical temperature ($T_c^D$) to the intra-valley critical temperature ($T^S_c$) as a function of $x$. b) Phase diagram for the even chirality C2DEG model of $J$-layer RG with $ J=4,6$ and $8$ in increasing order. The electron density $n_{e}$ is in units of cm$^{-2}$ and the Haldane mass is in meV. At low densities, $n_e < n_{e,2}$ the chiral-$J$ Kekul\'{e} superconducting channel dominates. c) Phase diagram for the odd chirality C2DEG model of $J$-layer RG with $ J=3,5$ and $7$ in increasing order, with the same axis label as in b). The chiral-$J$ Kekul\'{e} superconducting channel dominates for $n_e < n_{e,2}$, giving way to $\Qq =0$ valley triplet chiral superconductivity for $n_{e,1} < n_e < n_{e,2}$, with the phase boundary indicated by the solid line. Conventional valley singlet s-wave superconductivity dominates for $n_e >n_{e,2}$, with the dashed line indicating the phase boundary. d) Schematic of the Kekul\'{e} superconductor on the projected honeycomb lattice, with the superconducting unit cell indicated by the dotted lines. }
	\label{fig:LOchiralfig3}
	\end{center}
\end{figure}

For even values of $J$, the coherence factor $f_D(x)=(1-x^2)/2$ in Eq.~\ref{eq:RGinter}, arising from the quantum geometric properties of the Haldane phase, leads to the suppression of inter-valley pairing. The critical temperature for the intra-valley (S) pairing (with $f_{S}(x)=1$) is $T^{S}_c(x) = \omega_D \exp(-1/|g^{intra}_{+,l}|)$, higher than that of the inter-valley (D) pairing, $T^{D}_c(x) = \omega_D \exp(-2/((1-x^2)|g^{inter}_{+,l}|))$. As shown in Fig.~\ref{fig:LOchiralfig3} a), the ratio $T^{D}_c/T^{S}_c$  is exponentially suppressed, indicating that for even parity the intra-valley pairing dominates over the inter-valley pairing for all values of $x$. This exponential dependence is a direct result of the ratio of the even-parity inter- and intra-valley pair susceptibilities $\Pi_D/\Pi_S$ as derived in Eq.~\ref{eq:pairsusceptibility}.

Since intra-valley pairing dominates for even values of $J$ we focus on the interaction matrix elements associated with $G_{4}(\pp,\pp')$ and $G_{3}(\pp,\pp')$. For even chirality $J$, $(g_{+,l} > g_{-,l})$, with $ G_{+,0} = -G_0(1-x^2)$, $ G_{+,J} = -G_0 (1-x)/2$ and $G_{+,-J} = -G_0 (1+x)/2$. As $G_{+,-J} > G_{+,0} >G_{+,J}$ for $x_c > 0.5$ we find that the chiral-$J$ Kekul\'{e} superconductor  has the largest $T^S_c$ when $ k_F < (\sqrt{3}m/\zeta_J)^{1/J}$ corresponding to the densities $n_{e} < n_{e,2}$ with $n_{e,2} = 1/(2 \pi)(\sqrt{3}m/\zeta_J)^{2/J}$. The superconducting order parameter is given by $\Delta(\rr) = \Delta \cos(2 \Kk_D \cdot \rr + \theta)$  (where $\theta$ corresponds to the relative phase difference between the valleys). Based on the minimal chiral model, we obtain a phase diagram of superconductivity in Fig.~\ref{fig:LOchiralfig3} b) for even-parity chiral 2DEGs in the $(n_e,m)$ parameter space for different values of $J = 4, 6$ and $8$. To ensure a simply connected Fermi surface, we assume that the electron density $n_{e} \geq 0.8 \times 10^{12}$~cm$^{-2}$, sufficiently large~\cite{Han2025,Yoon2025a} to avoid the three-pocket and annulus Fermi surfaces induced by trigonal warping. As discussed in the End Matter, the superfluid stiffness of the chiral-$J$ Kekul\'{e} state scales as $n_e^{J/2}$. There, we also calculate the mean-field critical temperature of the chiral-$J$ Kekul\'{e} state in the range $T^S_c \sim 90.8 \textrm{mK}-13.5 \textrm{K}$.

%{\color{red} As detailed in the End Matter, the mean-field critical temperature of the chiral-$J$ Kekul\'{e} state can be of order $\sim 10 $K. However, since this is a two-dimensional system, the superconducting transition temperature $T_{BKT}$ is set by the phase (superfluid) stiffness rather than the pairing scale. To determine the Berezinskii–Kosterlitz–Thouless (BKT) transition temperature $T_{BKT}$, we use the Kosterlitz-Nelson criteria $T_{BKT} = \pi/2 \rho_s(T_{BKT})$ and the approximate relation $\rho_s(T) = \rho_s(1-(T/T_c)^4)$ for the temperature dependence of the superfluid stiffness. Assuming $T_{BKT} \ll T_c$, we get $T_{BKT} = \pi/2 \rho_s \sim $, where the superfluid stiffness is $\rho_s = n_{2D}/ 4m^{\star}$, where $m^{\star}$ is the effective mass. }

The superconducting phase diagram is quite different for odd-parity chiral 2DEGs, as the intra-valley pairing only dominates the inter-valley pairing when $ x > 1/\sqrt{3}$. Therefore, the chiral Kekul\'{e} superconducting order has the highest $T_c$, when $n_{e} < n_{e,1}$ with $n_{e,1} = 1/(2 \pi)(\sqrt{2}m/\zeta_J)^{2/J}$. Since $G_{-,-J} = - G_0/2(1+x)$ is the dominant channel for odd parity, the intra-valley pairing order exhibits an $\pi$-phase resulting in a Kekul\'{e} superconductor with the order parameter $\Delta(\rr) = \Delta \sin(2 \Kk_D \cdot \rr + \theta)$. $G_{-,J}$ remains the dominant channel for $1/2<x<1/\sqrt{3}$, resulting in a valley-triplet $\pi$-phase chiral-$J$ inter-valley superconducting order for the density regime $n_{e,1}<n_e<n_{e,2}$. However, when $n_e > n_{e,2}$, $G_{+,0} = - G_0 (1-x^2)$ becomes the dominant channel; we get a conventional valley singlet s-wave superconductor with $\Qq =0$. Fig.~\ref{fig:LOchiralfig3} c) shows the superconducting phase diagram for the odd-parity chiral 2DEGs as a function of $(n_e,m)$. 

Interestingly, for both parities, the presence of a valley degree of freedom permits even-order pairing channels, including a momentum-independent $s$-wave pairing when $\mu > m$. However, the $s$-wave pairing channel is suppressed as $\mu \to m$ given $  G_{+,0} = -G_0(1-x^2)$. This suppression originates from a combination of the Pauli exclusion principle and the quantum geometric properties of the Haldane phase. At the Dirac points $(\pm\Kk_D)$, the sublattice and valley degrees of freedom become coincident and act as a single pseudospin, thereby forbidding local (contact) interactions and rendering $s$-wave Kekul\'{e} order incompatible. Consequently, this enhances the relative stability of the chiral-$J$ Kekul\'{e} order, which emerges as the dominant channel in this regime. Moreover, due to the presence of a non-zero pair tunneling channel $G_{3,l}$, we expect the chiral-$J$ Kekul\'{e} state to be robust to the application of a displacement field, so long as $\Delta_V < m$. 

{\em Chiral Kekul\'{e} superconductivity}.---The proposed superconducting state exhibits a spatial Kekul\'{e} pattern on the projected bipartite lattice, indicated in Fig.~\ref{fig:LOchiralfig3} d, and notably, the size of the superconducting unit cell is tripled. The odd chiral-$J$ Kekul\'{e} superconductor is a valley triplet with the spontaneously broken symmetry $U(1) \otimes O_V(3) \otimes \mathbb{Z}_3$, where the $U(1)$ is associated with the overall superconducting phase, $O_V(3)$ is the valley symmetry, and $\mathbb{Z}_3$ is related to the lattice translational symmetry. Alternatively, the even chiral-$J$ Kekul\'{e} superconductor must be a valley singlet with the spontaneously broken symmetry $U(1) \otimes \mathbb{Z}_3$. 

Unlike the phase-modulated superconductors revealed in the quarter-metal regime~\cite{Han2025,Yoon2025a}, the Kekul\'{e}  superconductors in the half-metal regime are characterized by a spatial modulation of the amplitude of the superconducting order parameter (Fig.~\ref{fig:LOchiralfig3} d). Resolving the Kekul\'{e} superconductor would require atomic-scale measurements of the superconducting gap or related Bogoliubov quasiparticle features as a function of position on the lattice, which can be directly probed by scanning tunneling microscopy. At nanometer length scales, it would show a fully gapped spectrum associated with the Bogoliubov quasiparticles of the chiral superconducting state. This provides a clear distinction from a conventional Kekul\'{e} lattice distortion, which involves an actual modulation of the lattice~\cite{Nuckolls2023,Chen2024}.

%\changes{The Kekul\'{e} superconducting pattern can be directly probed by scanning tunneling microscopy using a superconducting tip. A normal metal tip would instead measure the quasiparticle density of states and detect a fully gapped spectrum, reflecting the Bogoliubov quasiparticle excitations of the chiral superconducting order parameter and thereby distinguishing it from a Kekul\'{e} lattice distortion~\cite{Nuckolls2023,Chen2024}.}

Kekul\'{e} superconducting textures have previously been proposed for Dirac fermions in graphene, arising from nearest-neighbor pairing interactions~\cite{PhysRevB.82.035429,PhysRevB.93.155149}. However, the mechanism we describe here is fundamentally different: it originates from topologically enforced geometric constraints of the Bloch wavefunctions, which enhances the pair susceptibility at large momenta ($\pm 2 \Kk_D$). As such, this Kekul\'{e} superconducting order can emerge from a broad class of attractive interactions in the Haldane phase of C2DEGs, independent of the detailed microscopic pairing mechanism. 

In RG, trigonal warping of the bands, due to remote hopping processes, induces Lifshitz transitions~\cite{PhysRevB.82.035409}, wherein the Fermi surface evolves from being circularly connected to an annular Fermi surface or a set of disconnected pockets. For tetralayer RG, this transition occurs in the half-metal regime at an electron density of $n_e \sim 0.4 \times 10^{12}$~cm$^{-2}$ near zero displacement field~\cite{Han2025,Yoon2025a}. Even above such a density, where the Fermi surface remains connected, it loses its azimuthal symmetry due to trigonal warping. In the presence of a trigonal warping energy scale $E_{tri}$, the weak-coupling enhancement of the intra-valley pair susceptibility is cut off by $E_{tri}$, and Eq.~\ref{eq:pairsusceptibility} retains the same functional form, but with $\log(E_c/T) \to \log (E_c/E_{tri})$. Consequently, for even chirality, a critical pairing strength $G_{+,-J} = -(1-x^2)^{\alpha}/(N_0 \log (E_c/E_{tri} )) $ is required to stabilize Kekul\'{e} superconductivity. 

Multiple superconducting phases have been detected in bilayer to hexalayer RG~\cite{Zhou2021a,Han2025,kumar2025superconductivitydualsurfacecarriersrhombohedral,morissette2025stripedsuperconductorrhombohedralhexalayer}
in the fully layer-polarized (FLP) states, induced by displacement fields. However, more recently, superconductivity has been observed in doped layer-antiferromagnetic (LAF) ground states near zero displacement field~\cite{kumar2025superconductivitydualsurfacecarriersrhombohedral}, notably, in the transition region between the FLP and LAF states, where the Haldane phase is expected to occur. In ultra-clean suspended bilayer graphene, an anomalous Hall state with $m \sim 1–3$ meV has been inferred experimentally~\cite{PhysRevLett.105.256806,PhysRevLett.108.076602}, while in hBN-encapsulated pentalayer RG devices a QAH state with $m \sim10–15$ meV has been reported~\cite{ABCrGSOC}. It would be valuable to investigate our proposal for superconducting states that arise at zero or small displacement fields in these regimes.

Assuming the Haldane phase is energetically favored, the remaining competition involves either a charge density wave or an inter-valley coherent phase, which leads to a Kekul\'{e} lattice distortion~\cite{Nuckolls2023,Chen2024}. Inter-valley coherence is typically favored only near charge neutrality ($\mu = 0$)~\cite{RevModPhys.84.1067}; however, in the presence of a Fermi surface ($\mu \neq 0$) the leading instability is toward superconductivity. Although the Kekul\'{e} distortion can be driven by optical phonons~\cite{Nuckolls2023,Chen2024}, our estimates in RG indicate that acoustic phonons play a stronger role. Charge-density-wave order is also unlikely, as it generally requires Fermi-surface nesting in the particle-hole channel, which is subleading in our case. Taken together, these considerations suggest chiral Kekul\'{e} superconducting state dominates at low densities for a connected Fermi surface in RG.

{\em Discussions and Outlook}.---In conventional metals, the emergence of PDWs typically requires strong correlation effects or fluctuations tied to lattice-scale physics, often leading to modulation periods incommensurate with the crystalline unit cell~\cite{PhysRevLett.88.117001,PhysRevB.81.020511,PhysRevX.4.031017,PhysRevB.89.165126,RevModPhys.87.457,PDWreview,PhysRevLett.125.167001,PDWDavis,PhysRevLett.99.127003,Agterberg2008,Berg2009,PhysRevB.91.104512,PhysRevB.97.174510,PhysRevB.97.174511,vortexhalosPDW,PhysRevLett.102.207004,PhysRevB.77.174502,QOcupratesPDW}. In contrast, we showed that in the Haldane phase of C2DEGs, the pair susceptibility is enhanced at finite momentum due to the topologically enforced quantum geometry of the Bloch wavefunctions. This enhancement occurs at high-symmetry points in the Brillouin zone, naturally favoring the formation of lattice-scale PDW states even in the weak-coupling regime. Unlike conventional scenarios that rely on strong-coupling mechanisms, competing orders, or fluctuation-driven instabilities, the quantum geometric mechanism proposed here provides a topologically protected mechanism to realize a lattice-scale PDW order in other chiral Chern bands of 2D crystals and layered materials.

%The experimental implications of lattice-scale PDWs, especially the proposed chiral Kekul\'{e} superconducting states, present an interesting direction for future studies.

\section{Acknowledgements}
Work by Y.B. was supported in part by the DOE EPSCoR program under the award DE-SC0022178, and a grant from NIST.  F.Z. was supported by the National Science Foundation under grants DMR-2414726, DMR-1945351, and DMR-2324033 and by the Welch Foundation under grant AT-2264-20250403. Work by E.R. was funded by the US Department of Energy, Office of Basic Energy Sciences, via Award DE-SC0022245. Two of us (Y.B. and E.R.) would like to recognize the Kavli Institute for Theoretical Physics (KITP) for hosting us as part of the moire24 program, which is supported by NSF PHY-2309135, where part of this research was performed.

%\bibliographystyle{apsrev4-2}
%\bibliography{ReferenceschiralKekule}

\begin{thebibliography}{101}
\bibitem [{foo({\natexlab{a}})}]{footnote1}%
  \BibitemOpen
  \href@noop {} {\bibinfo {title} {From here onward, we use the term {H}aldane
  phase to refer to the {QAH} phase induced by the mass term $m \hat{\sigma}_z
  \tau_z$ in {C2DEGs}, even though the microscopic mechanism leading to this
  phase can vary depending on the chirality index {$J$}}}
\BibitemShut {NoStop}%
\bibitem [{foo({\natexlab{c}})}]{footnote2}%
  \BibitemOpen
  \href@noop {} {\bibinfo {title} {We adopt a conservative estimate for the
  deformation field, noting that it can be up to twice as
  large~\cite{PhysRevLett.127.187001,PhysRevB.99.165112}. {A} detailed study is
  required to determine different electron–phonon scattering processes that
  can contribute to the magnitude of the deformation field for each {$J$}-layer
  {RG} system}} \BibitemShut {NoStop}%
\end{thebibliography}


\clearpage

\newpage

\begin{center}
\textbf{End Matter}
\end{center}

{\em Interaction effects in J-layer RG}.---For long-range interactions that are both valley- and spin-independent, the spontaneously broken symmetry quantum valley Hall (QVH) and quantum anomalous Hall (QAH) phases are energetically degenerate. To distinguish between these competing states, one must consider interactions that break valley symmetry or differentiate between the sublattices. Since both QVH and QAH mass terms share the same sublattice structure but differ in their valley dependence, only valley-dependent interactions can lift their degeneracy. Therefore, we only include valley-dependent interactions in our analysis. 

The interacting Hamiltonian, $H_{int} = H_{ee} + H'_{int}$, includes the SU(4) Coulomb interaction $H_{ee}$, with the Coulomb repulsion $v_{\qq}$, and the valley-dependent interaction $H'_{int}$
\bea 
\label{eq:Haminteracting}
\nonumber
H_{int} &=& \frac{1}{2L^2}\sum_{\kk,\kk' ,\qq}  v_{\qq} \psi^{\dagger}_{\kk+\qq}  \psi_{\kk} \psi^{\dagger}_{\kk'-\qq}  \psi_{\kk'} \\
&+& \frac{1}{2L^2}\sum_{\kk,\kk' ,\qq} \sum_{\mu =z,x,y} u_{\mu} \psi^{\dagger}_{\kk+\qq} \hat{\tau}_{\mu} \psi_{\kk} \psi^{\dagger}_{\kk'-\qq}  \hat{\tau}_{\mu} \psi_{\kk'},
\eea
where $u_{\mu}$'s are short ranged valley-dependent interaction terms. $\hat{\tau}_z$ acts on the valley degree of freedom and $L^2$ denotes the sample area. For simplicity, we assume that $u_{x} = u_{y} = u_{\perp}$.

\begin{figure}[b]
 \begin{center}
    \includegraphics[width=0.5\textwidth,clip]{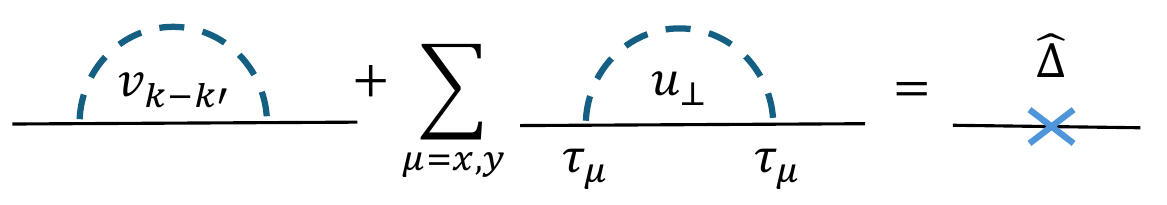}
\caption{Feynman diagram corresponding to the self-consistent solution to the gap equation for the mean field $\hat{\Delta}$.}
\label{fig:Feyntree}
\end{center}
\end{figure}

 To perform the mean-field theory, we rewrite the Hamiltonian $H$ by adding and subtracting the mean field terms $H_{\hat{\Delta}}$ as $H = (H_0 + H_{\hat{\Delta}}) + (H_{int} - H_{\hat{\Delta}}) = H_{MF} + H_{res} $, and $H_0$ corresponds to the non-interacting Hamiltonian. The mean-field Hamiltonian $H_{MF}$ is treated as the non-perturbative Hamiltonian, and the effect of $H_{res}$ can be studied via a diagrammatic approach. To ensure the field $ \hat{\Delta}$ is self-consistently determined, we satisfy the tree-level self-consistent equation corresponding to the Feynman diagrams shown in Fig.~\ref{fig:Feyntree}. We consider the competition between interaction-induced mass terms, one corresponding to the QAH effect, $m$, and the other to the QVH effect, $\Delta$. The QVH  Hamiltonian $H_{QVH} =  \sum_{\kk} \psi^{\dagger}_{\kk} \Delta_{VH} \hat{\sigma}_{z} \psi_{\kk} $, and the QAH Hamiltonian as
$H_{QAH} = \sum_{\kk} \psi^{\dagger}_{\kk} m \hat{\sigma}_{z} \hat{\tau}_z \psi_{\kk} $
where $\hat{\sigma}_z$ acts on the sublattice degree of freedom. In the presence of valley anisotropy, the self-consistent gap equation at the tree level becomes
\bea 
\nonumber
\hat{\Delta} &=& -\frac{1}{\beta} \sum_{n} \int \frac{d^2 k'}{(2 \pi)^2} \bigg[ v_{\kk - \kk'}  G_{MF} (\kk',\imath \omega'_n ) \\ 
&+& \sum_{\mu=x,y,z} u_{\mu} \hat{\tau}_{\mu}  G_{MF} (\kk',\imath \omega'_n ) \hat{\tau}_{\mu} \bigg],
\eea 
where $G_{MF}(\kk) = (\imath \omega_n - H_{MF}(\kk))^{-1}$ is the Green's function associated with the mean-field Hamiltonian, $\omega_n = (2 n +1 )\pi/\beta$ and $\beta =1/(k_{\beta}T)$. Performing the summation over the frequencies, we arrive at the gap equations:
\bea
\label{eq:MFHaldane}
m(\kk) &=& \int \frac{d^2 k'}{(2 \pi)^2}  \bigg[ v_{\kk-\kk'} - 2 u_{\perp} \bigg] \frac{m(\kk')}{2 E_{\kk'}} \tanh\bigg( \frac{\beta E_{\kk'}}{2} \bigg), \\ \nonumber
\Delta_{VH}(\kk) &=& \int \frac{d^2 k'}{(2 \pi)^2}  \bigg[ v_{\kk-\kk'} + 2 u_{\perp} \bigg] \frac{\Delta_{VH}(\kk')}{2 E_{\kk'}} \tanh \bigg( \frac{\beta E_{\kk'}}{2} \bigg)  ,
\eea  
where $ E^2_{\kk} = |\epsilon_{J,\kk}|^2 + m_{\kk}^2 $ for the QAH state, and $ E^2_{\kk} = |\epsilon_{J,\kk}|^2 + \Delta_{VH}(\kk)^2 $ for the QVH state~\cite{footnote3}. From the structure of Eqs.~\ref{eq:MFHaldane}, it is clear that only $u_{\perp}$ lifts the degeneracy of the QAH and QVH state; therefore, we take $u_{z} =0$. The above equations imply that the QAH is preferred over the QVH state for $u_{\perp} < 0$. The negative sign that distinguishes the QAH state from the QVH state is due to the anti-commutation of the Pauli matrices in the valley space $\hat{\tau}_z \hat{\tau}_x = - \hat{\tau}_x \hat{\tau}_z$ and  $\hat{\tau}_z \hat{\tau}_y = - \hat{\tau}_y \hat{\tau}_z$. 

\begin{figure}[h]
 \begin{center}
    \includegraphics[width=0.48\textwidth,clip]{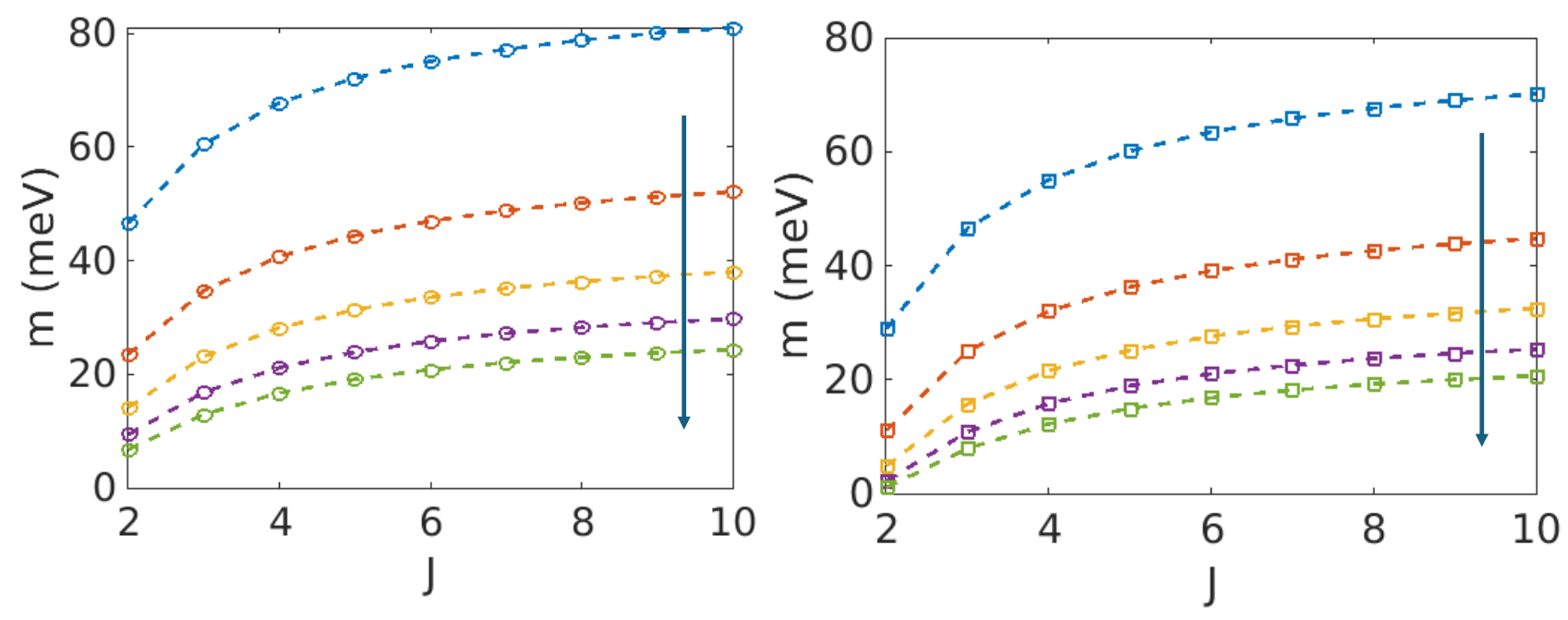}
\caption{Interaction induced mass gaps for different chirality indices $J$ for a) long-range Coulomb interactions and b) gate-screened interaction with $d =10$ nm evaluated at the Dirac point with a cutoff $k_c = \gamma_1/(\hbar v)$. Each curve corresponds to a different value of the dielectric constant $\epsilon = 5, 7.5, 10, 12.5$ and $ 15$, from the top to bottom, with an arrow indicating the direction of increasing $\epsilon$.}
\label{fig:Haldanemassgaps}
\end{center}
\end{figure}

Fig.~\ref{fig:Haldanemassgaps} a) shows the interaction-induced mass gaps as a function of the chirality index $
J$ for the bare Coulomb interaction $v_{\qq} = 2 \pi e^2/(\epsilon q)$ and $u_{\perp} < 0$ ($|u_{\perp}| \sim 2$ meV) computed for different values of the dielectric constant $\epsilon = 5, 7.5, 10, 12.5$ and $ 15$. In the self-consistent calculation, we impose a momentum cutoff $k_c = \gamma_1/v$ and evaluate $m$ at the Dirac point $\kk=0$. For low dielectric screening (e.g., $\epsilon = 5$), the interaction-induced mass gaps are substantial, with $m > 50$ meV. For reference, the dielectric constant of h-BN is approximately $\epsilon  \sim 6$. As expected, increasing the dielectric screening reduces the magnitude of the induced gaps. To account for the influence of nearby metallic gates, we incorporate a gate-screened Coulomb interaction $v^{sc}_{\qq} = v_{\qq} \tanh(qd/2)$ where $d$ is the distance between the gates. The results are plotted in Fig.~\ref{fig:Haldanemassgaps} b). Even in the presence of gate screening, the mass gaps remain sizable for $\epsilon = 5$, particularly in tetra- and pentalayer graphene, where we find $m>50 meV$.  

{\em Pair susceptibility in other channels}---One can calculate the particle-particle susceptibility in the $\hat{\sigma}_z$ channel. Here, we restrict to even chiralities. Again estimating the leading logarithmic contribution one finds, $\Pi_{-S} = x^2 \Pi_{S} $ and $\Pi_{-D} = (1-x^2)/2 \Pi_{S}$, where $\Pi_{-S}(\Pi_{-D})$ are the inter- and intra-valley pair susceptibility in the $\sigma_z$-channel ($\Pi_S$ denotes the even $J$ intra-valley pair susceptibility). The intra-valley pair susceptibility exceeds the inter-valley counterpart for $x > 1/\sqrt{3}$, indicating that in the $\hat{\sigma}_z$ channel intra-valley Kekul\'{e} superconductivity should be favored when $\mu < \sqrt{3} m$. One can also evaluate the pair susceptibility in the $\hat{\sigma}_x$ and $\hat{\sigma}_y$ channels for the Haldane phase. Here, we find that the inter-valley susceptibility dominates over the intra-valley component. The channel dependence highlights the sensitivity of the pair susceptibility on the topological nature and the orbital character of the pairing interaction.

{\em Pair Susceptibility in a displacement field}.---When inversion symmetry is explicitly broken, for example, by applying a finite displacement field ($\Delta_V\neq0$), preserving the topological phase requires $\Delta_V < m$. Simultaneously, realizing two Fermi surfaces, where a distinction between inter- and intra-valley pairing remains meaningful, necessitates  $ \Delta_V + m < \mu$. Defining $y= \Delta_V/\mu$, these constraints translate into the conditions $ y < 1-x$ and $y < x$. In this regime, the intra-valley pair susceptibility becomes valley-dependent and is given by 
\be
\Pi_{S}(\pm 2 \Kk_D) = \frac{N_0}{(1-(x \pm y)^2)^{\alpha}} \log\bigg( \frac{E_c}{ T} \bigg)f_{S}(x\pm y) + \cdots.
\ee
Meanwhile, the inter-valley pair susceptibility loses its weak-coupling logarithmic divergence due to the breaking of inversion symmetry. Finally, in the limit $y > 1-x$ and $y < x$, the system remains in the topological phase but hosts only a single Fermi surface at a single valley. This regime corresponds to a ``quarter-metal'' superconducting phase~\cite{Yoon2025a,wy3f-hgr9}.

\begin{figure}[t!]
 \begin{center}
    \includegraphics[width=0.49\textwidth,clip]{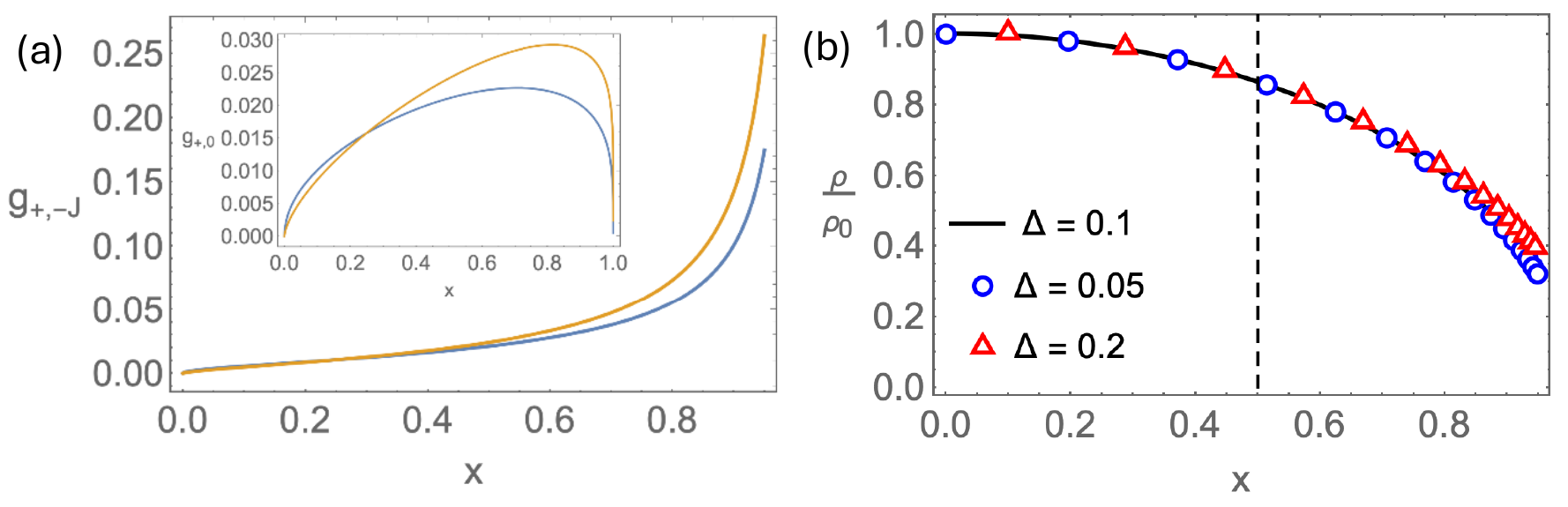}
\caption{(a) $g_{+,J}$ as a function of $x$ for a Haldane mass $m=10$ meV, for $J = 4$ and $6$ in increasing order, with the inset corresponding to $g_{+,0}$. (b) Ratio of the superfluid stiffness $\rho/\rho_0$ as a function of $x$ for a Haldane mass for $J = 4$ for different values of the order parameter $\Delta = 0.05, 0.1,$ and $0.2$ measured in unit of $\mu_0=(\mu^2-m^2)^{1/2}$. The vertical dashed line denotes the phase transition from the s-wave Kekul\'{e} to the chiral-J Kelul\'{e} superconductor at $x =1/2$.}
\label{fig:chiralKekEMFig3}
\end{center}
\end{figure}

{\em Electron repulsion}.---To account for the effect of electron repulsion, we project the
nearest-neighbor repulsive interaction~\cite{PhysRevLett.106.236805} onto the Fermi surface (see Supplemental Materials~\cite{supplement4}). Our calculations show that the intra-valley nearest-neighbor projected interaction contains only an $s$-wave component. Therefore, for even chirality, the chiral-$J$ Kekul\'{e} pairing interaction is unaffected by electron repulsion. In contrast, since the $s$-wave Kekul\'{e} state is influenced by the repulsive interaction, the corresponding phase boundary in Fig.~\ref{fig:LOchiralfig3} (b) shifts toward higher carrier densities, with the phase transition depending on the magnitude of the repulsive interaction. A similar argument indicates that the phase boundary of the odd-chirality Kekul\'{e} superconductor will also shift to higher densities. Electron repulsion beyond nearest-neighbor interactions will also affect the critical temperature by reducing the effective pairing interaction. A quantitative treatment of this effect requires an Eliashberg-type analysis, which is beyond the scope of the present work.

{\em Critical Temperature scaling}.---The critical temperature $T_c(x)$ for the even parity chiral-$J$ Kekul\'{e} superconductor is determined by $g_{+,-J}$, whereas the s-wave Kekul\'{e} superconductor is determined by $g_{+,0}$. Fig.~\ref{fig:chiralKekEMFig3} (a) shows the behavior of $g_{+,-J}(g_{+,0})$ as a function of $x$ for $J = 4$ and $6$, with a Haldane mass $m=10$ meV. Here, we use the effective pairing interaction $G_{0} =123 $ meV nm$^{2}$ which ignores electron repulsion. As $x \to 1$, $g_{+,-J}$ increases sharply due to the large density of states at the bottom of the band, while $g_{+,0} \to 0$ due to suppression of the s-wave channel. At $x=0.5$, we find $g_{0,-J} >g_{+,0}$, resulting in the s-wave Kekul\'{e} superconducting phase at higher densities. Using $ \omega_{D} \sim 0.2 $ eV and $g_{+,-J} \sim 0.1-0.2$, for $J =4,6$, yields $T^S_c \sim 90.8 \textrm{mK}-13.5 \textrm{K}$ for the chiral-$J$ Kekul\'{e} superconductor.

{\em Superfluid Stiffness}.---The superfluid stiffness of chiral 2DEGs, $\rho(m)$, contains both conventional and quantum-geometric contributions~\cite{PhysRevB.109.214518}. In the weak coupling limit, the stiffness is dominated by the conventional contribution, which is independent of $\Delta$ (see Supplemental Materials~\cite{supplement5})~\cite{fw3r-pcw5}. For $m \neq 0$ and $\Delta/\mu_0<0.2$, where $\mu_0 = \sqrt{\mu^2 - m^2}$, we find
\begin{equation}
\label{eq:ChStiffness}
\rho(x)=\frac{J\zeta_J (2 \pi n_{e})^{J/2}}{\pi} h(x) = \rho_0 h(x),
\end{equation}
where $h(x)$ is a scaling function independent of both $J$ and $\Delta$. The geometric stiffness $\rho_g$ scales as $\rho_g/\rho_0 \sim \Delta^2/(\mu^2-m^2)$, therefore, in the weak-coupling limit it becomes negligible. This zero-temperature expression for stiffness is valid for both inter- and intra-valley superconducting states and for any symmetry of the order parameter. 

For the Kekul\'{e} superconducting phase, $\rho(x)$ is shown as a function of $x$ in Fig.~\ref{fig:chiralKekEMFig3} (b).  The results show that the stiffness is independent of the value of $\Delta$ in the weak coupling limit. In the Kekul\'{e} superconducting state, which occurs at $x=1/2$, the stiffness is approximately $\rho \approx 0.8\rho_0$. As $x$ increases, the stiffness decreases to $\rho \approx 0.4 \rho_0$ for $x \sim 0.95$, but it is still determined by the electron density $n_{e}$. As in a conventional two-dimensional electron gas, where Galilean invariance gives $\rho_s=\hbar^2 n_{e}/m^\star$, the superfluid stiffness in a chiral 2DEG with $\Delta\ll\mu$ is governed primarily by the carrier density $n_e^{J/2}$, with only weak dependence on the magnitude of the superconducting gap.

\end{document}